\documentclass{PoS}
\RequirePackage{xspace}
\def\bbbar {\ensuremath{b \overline b}\xspace}
\def\to {\ensuremath {\rightarrow}\xspace}
\def\xbb {\ensuremath {X_{b \overline b}}\xspace}

\title{Exclusive Study of Bottomonium States in Radiative $\Upsilon(2S)$ Decays}

\ShortTitle{Bottomonium States in Radiative $\Upsilon(2S)$ Decays}

\author{\speaker{S.~Sandilya}\thanks{On behalf of Belle Collaboration}\\
  Tata Institute of Fundamental Research, Mumbai 400005\\
  E-mail: \email{saurabh@tifr.res.in}}

\abstract{A study of various bottomonium states, in $\Upsilon(2S)\to\gamma (\bbbar)$ decays, reconstructed exclusively in 26 hadronic final states is presented. The study is performed using a data sample recorded at the $\Upsilon(2S)$ resonance with the Belle detector at KEKB, that contains $157.8\times10^6$ $\Upsilon(2S)$ events. The $\chi_{bJ}(1P)$ states are found with their masses being consistent with the world average values. We find no evidence for the state claimed to have been observed around 9975~$\rm MeV/c^{2}$ in an analysis based on a data sample of $9.3\times 10^6$ $\Upsilon(2S)$ events collected with the CLEO~III detector, and place an upper limit an order of magnitude lower than the latter result. In the same study, the $\eta_{b}(1S)$ state is also searched for.}

\FullConference{XV International Conference on Hadron Spectroscopy-Hadron 2013\\
  4-8 November 2013\\
  Nara, Japan }

\begin{document}
\section{Introduction}
\label{intro}
\emph{Bottomonium}, a bound system of a $b$ quark and an anti-$b$ ($\bar{b}$) quark, is the heaviest known quarkonium. The bottomonium system is almost non-relativistic $(v^{2}/c^{2} \approx 0.1)$, hence it offers a unique laboratory to study spin-dependent QCD interactions and to test QCD calculations~\cite{Brambilla:2012cs}. In these proceedings, the discussion is focused on the recent refutation of the $X_{b \bar b}(9975)$ state~\cite{Sandilya:2013rhy} while also covering the $\eta_{b}(1S)$, $\eta_{b}(2S)$, and $\chi_{bJ}(1P)$ states.

\section{\boldmath The $\eta_{b}(2S)$ and $X_{b \bar b}(9975)$ states}
\label{xbb}
First evidence of the $\eta_{b}(2S)$ state was reported by the Belle collaboration in the radiative transition $h_{b}(2P)\to\eta_{b}(2S)\gamma$, with a significance (including systematics) of $4.2\sigma$~\cite{Mizuk:2012pb}. The analysis used a 133.4 fb$^{-1}$ data sample collected near the $\Upsilon(5S)$ resonance. It studied the process $e^{+}e^{-}\to\Upsilon(5S)\to h_{b}(nP)\pi^{+}\pi^{-}$, $h_{b}(nP)\to\eta_{b}(mS)\gamma$, in which the $\eta_{b}(mS)$ states are reconstructed inclusively. It reported the mass of the $\eta_{b}(2S)$ state to be $9999.0\pm3.5^{+2.8}_{-1.9}$ $\rm MeV/c^{2}$, which corresponds to a hyperfine mass splitting between the $\Upsilon(2S)$ and $\eta_{b}(2S)$ states, $\Delta M_{\rm HF}(2S)\equiv M[\Upsilon(2S)]- M[\eta_{b}(2S)]$, of $24.3^{+4.0}_{-4.5}$ $\rm MeV/c^{2}$.

Recently, observation of a bottomonium state, $\xbb$, has been made in the radiative $\Upsilon(2S)$ decay, in an analysis performed on a data sample of $9.3 \times 10^{6}$  $\Upsilon(2S)$ decays recorded with the CLEO III detector~\cite{Dobbs:2012zn}. The $\xbb$ state is reconstructed in 26 exclusive hadronic final states, and it has been observed with a significance of $\sim 5 \sigma$ at a mass of $9974.6 \pm 2.3 \pm 2.1$ $\rm MeV/c^{2}$. In that paper, the $\xbb$ state is assigned to the $\eta_{b}(2S)$ state, which corresponds to the hyperfine splitting, $\Delta M_{\rm HF}(2S)= 48.6\pm3.1$ $\rm MeV/c^{2}$ . The claim of $\xbb$ state to be the $\eta_{b}(2S)$ is in disagreement with the Belle result in Ref.~\cite{Mizuk:2012pb} as well as with the lattice QCD, potential model and related theoretical predictions that are summarized in Ref.~\cite{Burns:2012pc}. On the other hand, the Belle result~\cite{Mizuk:2012pb} is in agreement with theory~\cite{Burns:2012pc,Dowdall:2013}.

With almost 17 times larger data (at the $\Upsilon(2S)$ resonance) than the CLEO-III, Belle had a unique opportunity to confirm or refute the aforementioned observation of the $\xbb$ state. Belle performed an analysis to search the $\xbb$ state in the reaction, $\Upsilon(2S) \to \gamma \xbb $~\cite{Sandilya:2013rhy}, where the $\xbb$ state is reconstructed in the same 26 modes as mentioned in Ref.~\cite{Dobbs:2012zn}. The study is done on the 25 fb$^{-1}$ data sample collected at the $\Upsilon(2S)$ resonance (containing $157.8 \times 10^{6}$ $\Upsilon(2S)$ decays~\cite{NY2S}) by the Belle detector~\cite{belle} at the KEKB asymmetric-energy $e^+e^-$ collider~\cite{KEKB}. In addition, off-resonance data samples, 89.5 fb$^{-1}$ collected 60 MeV below the $\Upsilon(4S)$ resonance and 1.7 fb$^{-1}$ collected 30 MeV below the $\Upsilon(2S)$ resonance, are used for the study of the continuum background.

Charged tracks are selected to form a  $b\bar{b}$ system, imposing the impact parameter requirements, and are later identified as kaons, pions and protons based on information from particle identification subdetectors (central drift chamber, time of flight and aerogel Cherenkov counters). An isolated (not matched with a charged track) cluster in the barrel electromagnetic calorimeter that has energy greater than 22 MeV and cluster shape consistent with an electromagnetic shower is selected as a photon candidate. An $\Upsilon(2S)$ candidate is formed by combining a photon candidate with the $b\bar{b}$ system. To suppress continuum background, a requirement on $|\cos\theta_{T}|<0.8$ is imposed, where $\theta_{T}$ is the angle between the photon candidate and the thrust axis in the event. Later, requirements on the difference between the energy of the $\Upsilon(2S)$ candidate and the beam energy in the center-of-mass (CM) frame ($- 0.04$ GeV $< \Delta E < $ 0.05 GeV), the momentum of the $\Upsilon(2S)$ candidate in the CM frame ($P^{*}_{\Upsilon(2S)} < $ 0.03 GeV/c), and the angle between the $\gamma$ candidate and the $b\bar{b}$ system in the $\Upsilon(2S)$ candidate CM frame ($\theta_{\gamma b\overline b} >$ $150^{\circ}$). These requirements are obtained from an optimization procedure by using $S/\sqrt{S+B}$ as a figure of merit, where \emph{S} is the number of signal events estimated from the branching fraction obtained by Ref.~\cite{Dobbs:2012zn} and \emph{B} is the number of background events. A kinematic fit (4C-fit) is applied to the $\Upsilon(2S)$ candidates, which improves the resolution of the signal, and the $\chi^{2}$ of the 4C-fit is used to select the best candidate.

\begin{figure}[htbp]
\includegraphics[width=15.2cm]{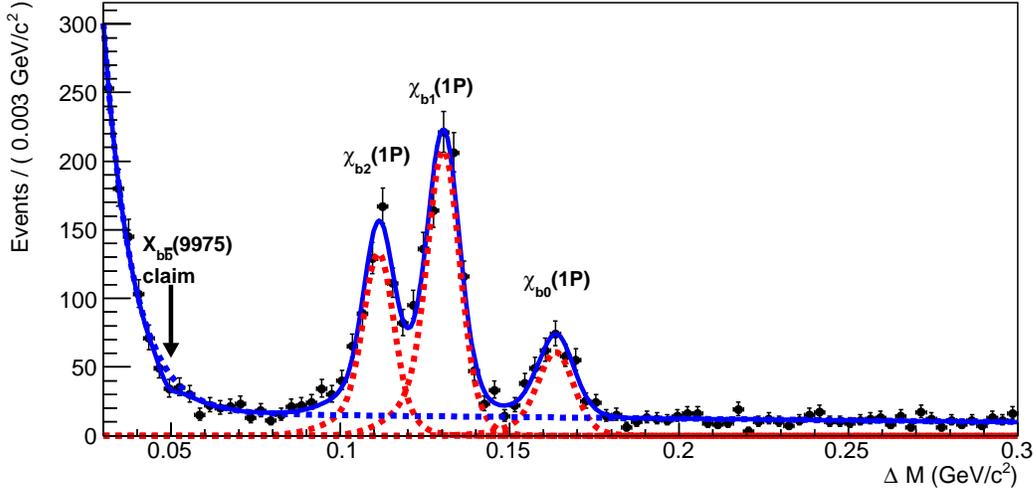}
\caption{$\Delta M$ distributions for $\Upsilon(2S)$ data events that pass the selection criteria applied. Points with error bars are the data, and the blue solid curve is the result of the fit for the signal-plus-background hypothesis, and the blue dashed curve is the background component. The three $\chi_{bJ}(1P)$ components indicated by the red dotted curves are considered here as part of the signal.}
\label{fig:etab2s}
\end{figure}

In Figure~\ref{fig:etab2s}, the  $\Upsilon(2S)$ data after all selection criteria applied is presented in terms of $\Delta M \equiv M[(b\overline b) \gamma]-M[(b\overline b)]$. The signals [$\xbb$ and $\chi_{bJ}(1P)$] in the plot are parameterized by the double gaussian (a standard gaussian and an asymmetric gaussian) and the background is parameterized by a sum of an exponential function and a first order Chebyshev polynomial (shown with blue dotted line in Fig.~\ref{fig:etab2s}). The large signal yields for the $\chi_{bJ}(1P)$ are found (300, 950, and 580 events for J=0, 1, and 2 respectively), which allows to precisely determine mass of $\chi_{bJ}(1P)$ states in agreement with PDG 2012~\cite{pdg2012} to be $9859.63 \pm 0.49$, $9892.83 \pm 0.23$, and $9912.00 \pm 0.34$$ \rm MeV/c^{2}$, respectively, for \emph{J} = 0, 1, and 2. No signal ($-30 \pm 19$ events) is found for the \xbb state and an upper limit on the product branching fraction ${\cal B}[\Upsilon(2S)\to\xbb\gamma]\times\sum_i{\cal B}[\xbb\to h_i]$ $< 4.9\times 10^{-6}$ is derived at 90\% CL (Confidence Level).

\section{\boldmath The $\eta_{b}(1S)$ state}
The ground state of bottomonium system, the $\eta_{b}(1S)$ meson, was first observed in the radiative decays of $\Upsilon(3S)$ by the BaBar collaboration in 2008~\cite{Aubert:2008ba}. The measurements done for the $\eta_{b}(1S)$ state, are summarized in Table~\ref{tab:m_etab1}. The Belle result is not only the single most precise measurement of the $\eta_{b}(1S)$, but also it measures the width of $\eta_{b}(1S)$ to be $10.8^{{+4.0}{+4.5}}_{{-3.7}{-2.0}}$ MeV for the first time~\cite{Mizuk:2012pb}.
\begin{table}
  \caption{Measurements for the $\eta_{b}(1S)$ mass.}
  \begin{center}
  \begin{tabular}{c|c|c}
    \hline
    Reference & Transition & Mass of $\eta_{b}(1S)$ in $\rm MeV/c^{2}$  \\
    \hline
    Belle Collab.~\cite{Mizuk:2012pb}&$e^{+}e^{-}\to\Upsilon(5S)\to h_{b}(nP)\pi^{+}\pi^{-}$, & $9402.4 \pm 1.5 \pm 1.8$ \\
    &$h_{b}(nP)\to\eta_{b}(1S)\gamma$, where n = 1 and 2 & \\
    CLEO Collab.~\cite{Bonvicini:2009hs}& $\Upsilon(3S)\to\gamma\eta_{b}(1S)$ & $9391.8 \pm 6.6 \pm 2.0$ \\ 
    BaBar Collab.~\cite{Aubert:2009as}&$\Upsilon(2S)\to\gamma\eta_{b}(1S)$ & $9394.2^{+4.8}_{-4.9}\pm 2.0$\\
    BaBar Collab.~\cite{Aubert:2008ba}&$\Upsilon(3S)\to\gamma\eta_{b}(1S)$ & $9388.9^{+3.1}_{-2.3}\pm 2.7$ \\
  \end{tabular}
  \end{center}
  \label{tab:m_etab1}
\end{table}

\begin{figure}[htbp]
  \centering
  \includegraphics[width=8.0cm]{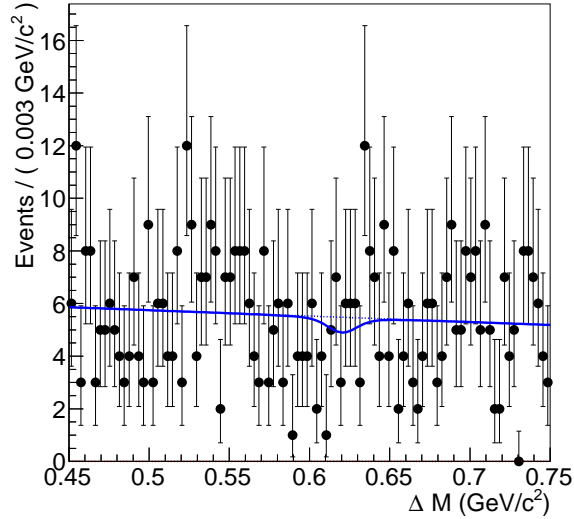}
  \caption{$\Delta M$ distributions for $\Upsilon(2S)$ data events that pass the selection criteria applied. Points with error bars are the data and the blue solid curve is the result of the fit for the signal-plus-background hypothesis, and the blue dashed curve is the background component.}
  \label{fig:etab1s}
\end{figure}

Similar to the search of the \xbb state (as discussed in Section~\ref{xbb}), the $\eta_{b}(1S)$ state is also looked for in the same 26 exclusive hadronic states in the radiative decays of $\Upsilon(2S)$~\cite{Sandilya:2013rhy}. Figure ~\ref{fig:etab1s} shows the $\Delta M$ distribution for the selected $\Upsilon(2S)$ data events, where no signal of the $\eta_{b}(1S)$ state is found in the exclusive reconstruction.  In that paper, Belle reported an upper limit at 90\%CL on the product branching fraction ${\cal B}[\Upsilon(2S)\to\eta_{b}(1S)\gamma]\times\sum_i{\cal B}[\eta_{b}(1S)\to h_i]$ $< 3.7\times 10^{-6}$. 

\section{Summary}
Belle performed an analysis in the search of the $\xbb$ state in the radiative $\Upsilon(2S)$ decay, reconstructed in the same 26 hadronic modes in Ref.~\cite{Dobbs:2012zn}, but no enhancement seen near 9975 $\rm MeV/c^{2}$. So, an upper limit on the product branching fraction ${\cal B}[\Upsilon(2S)\to\xbb\gamma]\times\sum_i{\cal B}[\xbb\to h_i]$ $< 4.9\times 10^{-6}$ is derived at 90\% CL. In case of $\eta_b(1S)$ reconstructed exclusively in the 26 modes, Belle gave an upper limit on the product branching fraction ${\cal B}[\Upsilon(2S)\to\eta_{b}(1S)\gamma]\times\sum_i{\cal B}[\eta_{b}(1S)\to h_i]$ $< 3.7\times 10^{-6}$ is derived at 90\% CL.

\acknowledgments
I am thankful to the organizers of the {\emph {XV International Conference on Hadron Spectroscopy}} for giving me an opportunity to present the talk and also providing me partial financial support to attend the conference. I am grateful to S.~Eidelman, R.~Mizuk, G.~B.~Mohanty, R.~Mussa, M.~Nakao, Y.~Sakai, C.~P.~Shen, K.~Trabelsi and S.~Uehara for their helpful suggestions during the preparation of the talk and these 
proceedings.

\end{document}